\documentclass[preprint,showpacs,preprintnumbers,amsmath,amssymb,aip,jcp]{revtex4}
\usepackage[dvipdfmx]{graphicx}
\usepackage{dcolumn}% Align table columns on decimal point
\usepackage{bm}
\begin{document}
\title{Coil--globule transition of a polymer involved in excluded-volume interactions with macromolecules 
}
%\title{Free energy of the coil--globule transition of polymers in a binary mixture}
%
\author{Kenta Odagiri}
%\altaffiliation[Permanent address:]{
\affiliation{
School of Network and Information, Senshu University, Kawasaki 214-8580, Japan}
\author{Kazuhiko Seki}
\email{k-seki@aist.go.jp}
\affiliation{
National Institute of Advanced Industrial Science and Technology (AIST)\\
AIST Tsukuba Central 5, Tsukuba, Ibaraki 305-8565, Japan}
%\email{
%
%}

%\date{\today}
\preprint{}
%%%%%%%%%%%%%%%%%%%%%%%%%%%%%%%%%%%%%%%%%%%%%%%%%%%%%%%%%%%%%%%%%%%%%%
% ABSTRACT  %%%%%%%%%%%%%%%%%%%%%%%%%%%%%%%%%%%%%%%%%%%%%%%%%%%%%%%%%%
%%%%%%%%%%%%%%%%%%%%%%%%%%%%%%%%%%%%%%%%%%%%%%%%%%%%%%%%%%%%%%%%%%%%%%
%\thispagestyle{empty}
\begin{abstract}
%\begin{center}
%Abstract less than 100 words.in single molecular level 
Polymers adopt extended coil and compact globule states  
according to the balance between entropy and interaction energies. 
The transition of a polymer between an extended coil state and compact globule state 
can be induced by changing thermodynamic force such as temperature to alter the energy/entropy balance. 
Previously, this transition 
was theoretically studied by taking into account 
the excluded-volume interaction between monomers of a polymer chain using the partition function. 
For binary mixtures of a long polymer and short polymers, 
the coil-globule transition can be induced by changing the concentration of the shorter polymers. 
Here we investigate 
the transition caused by short polymers by generalizing the partition function of the long polymer  to include the 
excluded-volume effect of short polymers. 
The coil-globule transition is studied as a function of the concentration of mixed polymers by 
systematically varying Flory's $\chi$-parameters. 
We show that the transition is caused by the interplay between the excluded-volume interaction and 
the dispersion state of short polymers in the solvent. 
We also reveal that the same results can be obtained by combining the mixing entropy and elastic energy 
if the volume of a long polymer is properly defined. 

%\end{center}
\end{abstract}
%\PACS 78.55.Qr \sep 72.20.Ee \sep  05.40.a

\maketitle
\newpage

\section{Introduction}
A single polymer molecule can transition between elongated coil and compact globule states
upon changing the temperature, pH of salts or 
addition of other polymer molecules. 
In general, the extended (coil) state of a long polymer is stabilized when 
the interaction between the polymer segment and solvent is favorable, typically occurring 
in good solvents. \cite{Sun} 
In contrast, the contracted (globule) state of a long polymer is stabilized when 
the interaction between the polymer segments is favorable, which occurs 
in poor solvents. 
By varying the solvent quality, the coil-globule transition of a polymer can be induced. 

Transitions between coil and globule states can also be induced by adding solutes that can be regarded as 
another solvent. \cite{Zhang,Sagle,Matsuyama,Budkov,Heyda}
When a good solvent is added to another good solvent, 
                                                                                                                                                                                                                                                                                                                                                                                                                                                                                                                                                                                                                                                                                                                                                                                                                                                                                                                                                                                                                                                                                                                                                                                                                                                                                                                                                                                                                                                                                                                                                                                                                                                                                                                                                                                                                                                                                                                                                                                                                                                                                                                                                                                                                                                                                                                                                                                                                                                                                                                                                                                                                                                                                                                                                                                                                                                                                                                                                                                                                                                                                                                                                                                                                                                                                                                                                                                                                                                                                                                                                                                                                                                                                                                                                                                                                                                                                                                                                                                                                                                                                                                                                                                                                                     a coil-to-globule transition may be observed as if the long polymer is in a homogeneous poor solvent. 
Similarly, an unintuitive transition from globule to coil can be obtained by mixing a polymer with poor solvents. 
The added solute can be short polymers. 
When short polymers are added, 
both coil-to-globule and globule-to-coil transitions have been observed. \cite{Maeda,Vasilevskaya,Kojima}
Inspired by the complex transitions of polymers induced by addition of polymer solvents, 
here we study theoretically the structural transformations of a long polymer induced by polymer solvents. 
%%% revised 
From another point of view, addition of polymer solvents puts a polymer in crowded  environment 
(macromolecular crowding).
Coil-globule transition can be regarded as the structural transition of a long polymer 
caused by ``macromolecular crowding".
%A major interaction in crowded environments originates from the excluded volume effect.   
%%% revised 

A qualitative description of the coil--globule transition was initially formulated using mean field theory. 
\cite{Flory,deGennes72,Rubinshtein}
Later, the nature of this phase transition, that is, whether it was continuous or discontinuous one, was extensively 
investigated using simulations, self-consistent field theory and sophisticated models that considered polymer stiffness. \cite{Birshtein,Moore,Grosberg,Grosberg70,Muthukumar,Maffi}
However, the additional effect of short polymers on the coil--globule transition of long polymers
has not yet been fully studied. %\cite{Maeda,Vasilevskaya,Kojima}
Part of the observed concentration dependences has been reproduced by 
phenomenological mean field theories, but the results were not
further justified using statistical theories. \cite{Maeda,Vasilevskaya,Kojima}
Although the coil--globule transition of polymers has been widely studied both theoretically and experimentally, 
it remains difficult to construct theories when a single molecular transition is induced by a change in the concentration of other polymers.

Here we study the transition between coil and globule states of long polymers induced by addition of short polymers 
by a statistical theory using the  partition function. 
The excluded-volume effect of short polymers is taken into account as a factor that decreases the emptiness in the partition function in a mean field sense. 
We also include Flory's $\chi$-parameters to take into account the interactions between the long polymer, short polymers and solvents. 
From the extremum of the partition function, we obtain an optimal expansion factor for a long polymer.

Among many possible conformations, 
the optimal ones can be determined from both the entropy in the long polymer
and its mixing entropy with other polymer molecules as well as their interaction energy. 
There is some ambiguity in combining the mixing entropy of polymers and elastic energy originating from entropy in a single polymer molecule.  
Using the partition function, such ambiguity is removed. We show that 
the same equation for the expansion factor can be derived by combining the mixing entropy of polymers and the elastic energy originating from the entropy in a single polymer molecule. 

In our approach, the optimal expansion factors are obtained from the extrema of the partition function, whereas 
the steady-state expansion factor was obtained by imposing the osmotic pressure balance at the boundary of the region occupied by a long polymer   
in previous approaches. \cite{Maeda,Vasilevskaya,Kojima}
In these theories, the region occupied by a long polymer is tacitly assumed to be in a phase different from that of the outer region 
for short polymers and the solvent. 
Furthermore, in principle, the pressure balance should include the hydrostatic pressure. 
In these theories, osmotic pressure was taken into account, 
but the hydrostatic pressure was not fully described. 
As in the previous theories, we ignore the hydrostatic pressure. 
Our results, nevertheless, reproduce the classical results obtained by virial expansion in the absence of short polymers.

Various types of shape transformations of the long polymer can be obtained 
when the solvent acts as a good solvent for the long polymer and a poor solvent for short polymers and vice versa. 
These competitive interactions and the excluded-volume interaction cause the shape transformation of the long polymer.
By systematically changing Flory's $\chi$-parameter between the long polymer, solvent and short polymers, 
we investigate the shape transformation of the long polymer induced by adding short polymers.

This paper is organized as follows.
In Sec. \ref{sec:Partition function}, the shape equation determining the optimal values of the expansion factor is 
derived from the partition function.
We also present an alternative derivation of the shape equation from the mixing entropy
using the effective volume of the long polymer.
In Sec. \ref{sec:shapeeq}, we qualitatively examine the shape equation using an approximation. 
Numerical solutions of the shape equation are described in Sec. \ref{sec:Shape transitions}. 
In Sec. \ref{sec:equilibrium shape}, the chemical equilibrium between inside and outside of the region occupied by a long polymer is considered. 
Sec. \ref{sec:Discussion} presents conclusion.

%%%%%%%%%%%%%%%%%%%%%%%%%%%%%%%%%%%%%%%%%%%%%%%%%%%%%%%%%%%%%%%%%%%%%%
\section{Derivation of the shape equation}
\label{sec:Partition function}

In this section, we derive the shape equation of a long polymer from the partition function.
We study the coil--globule transition of a long polymer with a degree of polymerization of $m_c$
in the presence of other polymers
with a degree of polymerization of $m_p$.
For both types of polymers, we introduce the Kuhn length of the monomer denoted by $\ell$. \cite{Rubinshtein}
We consider the case where $m_c \gg m_p$.  
We focus on the coil--globule transition of the long polymer and do not study the shape transition of short polymers here. 
The excluded volume of monomers of the long polymer is taken into account in the partition function. 
The short polymers are regarded as ideal chains.

%%%%%%%%%%%%%%%%%%%%%%%%%%%%%%%%%%%%%%%%%%%%%%%%%%%%%%%%%%%%%%%
\begin{figure}
\centerline{
\includegraphics[width=0.5\columnwidth]{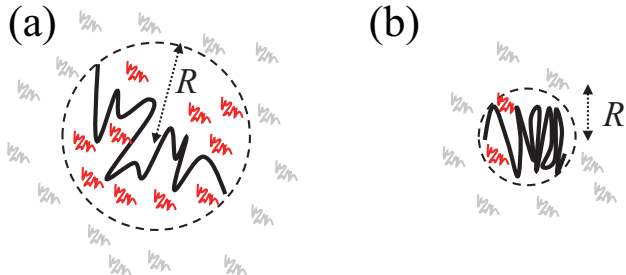}
}
\caption{
Schematic illustration of the long  polymer and short polymers.
(a) Elongated coil state, and (b) compact globule state.
$R$ denotes the gyration radius of the long polymer, and
dashed circles indicate the region occupied by the long polymer.
Small red zigzags represent short polymers inside the region occupied by the long polymer.
Small gray zigzags represent short polymers outside the region occupied by the long polymer.
}
\label{fig:inside}
\end{figure}
%%%%%%%%%%%%%%%%%%%%%%%%%%%%%%%%%%%%%%%%%%%%%%%%%%%%%%%%%%%%%%%

\subsection{Derivation by partition function}
The region occupied by a long polymer is characterized by the gyration radius
$R$ (Fig. \ref{fig:inside}).
We also introduce the natural end-to-end distance of the long polymer denoted as $R_0$. 
$R_0=\ell \sqrt{2 m_c/3}$ can be obtained from the extremum of 
$\ln W_0(R)$, where $W_0(R)$ is given by \cite{Rubinshtein}
\begin{align}
W_0 (R) = R^2 \exp \left( - \frac{3R^2}{2 m_c \ell^2} \right) 
\end{align}
 for 
an ideal Gaussian chain. 
The expansion factor of the chain is given by $\alpha=R/R_0$. 
The volume fraction occupied by the long polymer inside the region within $R$ of the long polymer is expressed as, 
\begin{align}
\Phi = m_c \ell^3/R^3 .  
\label{eq:p_1}
\end{align}
$\Phi $ is expressed using the expansion ratio as 
\begin{align}
\Phi = \frac{s}{\sqrt{m_c}} \frac{1}{\alpha^3}, 
\label{eq:Phi}
\end{align}
where $s = (\frac{3}{2})^{\frac{3}{2}}$ is a numerical factor independent of $\alpha$ and $m_c$.

The short polymers can move into the region occupied by the long polymer, as depicted in Fig. \ref{fig:inside}. 
The number of short polymers inside the region occupied by a long polymer is denoted as $n_i$. 
The volume fraction of short polymers inside the region occupied by the long polymer is denoted as $m_p p_i$ and 
can be obtained from 
$m_p p_i= n_i m_p \ell^3/R^3$.

An expression
for the partition function in the absence of short polymers was described by Di Marzio. \cite{diMarzio,Shew,ShewYoshikawa} 
In the presence of short polymers, 
the partition function can be generalized to (see Appendix \ref{sec:AppA} for details)
\begin{multline}
W (R) = R^2 \exp \left( -  \frac{3R^2}{2 m_c\, \ell^2} \right) \times  \\
\frac{1}{n_i! }
\left[ \prod_{k=0}^{n_i-1} \left(\frac{R^3}{\ell^3}\right) \left(1-k m_p \frac{\ell^3}{R^3}  \right)^{m_p} \right]
\left[ \prod_{j=0}^{m_c-1} \left( 1- m_p p_i - j  \frac{\ell^3}{R^3} 
\right)
\right] \times 
 \\
\exp \left[ 
-m_c \chi_{\Phi,s} \left(1-m_p p_i -\Phi \right) - m_c \chi_{\Phi,p} m_p p_i 
- m_p n_i\chi_{p,s} \left(1-m_p p_i - \Phi \right) 
\right], 
\label{eq:p_2}
\end{multline}
where $\chi_{\Phi,p}$, $\chi_{\Phi,s}$, and $\chi_{p,s}$
represent Flory's $\chi-$parameter between the long and short polymers, the long polymer and solvents, and the short polymer and solvents, respectively. \cite{Flory}
Flory's $\chi-$parameters are given by the interaction energies divided by $k_{\rm B} T$, 
where $k_{\rm B}$ is the Boltzmann constant and $T$ is temperature. 
The occupancy probability is assumed to be proportional to the volume fraction of emptiness. \cite{diMarzio,Shew,ShewYoshikawa} 
The factor given by $\left(1-m_p p_i -j  \ell^3/R^3\right)$ represents the volume fraction of emptiness to place the 
$j+1$th segment of the long polymer when short polymers and $j$ segments of the long polymer are already placed. 
Similarly, the factor given by $\left(1-k m_p \ell^3/R^3\right)$ represents the emptiness to place 
the $k+1$th short polymer in the region occupied 
by the long polymer.

In Eq. (\ref{eq:p_2}), we rewrite the term relating the occupancy probability as 
\begin{align}
\nu(R)=\prod_{j=0}^{m_c-1} 
\left( 1- m_p p_i - j \frac{\ell^3}{R^3} 
\right)&=\prod_{j=0}^{m_c-1} 
\frac{\ell^3}{R^3} \left[\frac{R^3}{\ell^3}\left( 1- m_p p_i\right)-j \right] \\
&=\left(\frac{\ell}{R}\right)^{3 m_c}
\frac{\Gamma\left[\frac{R^3}{\ell^3}\left( 1- m_p p_i\right)+1\right]}{\Gamma\left[\frac{R^3}{\ell^3}\left( 1- m_p p_i\right)-m_c+1\right]}
, 
\label{eq:p_3}
\end{align}
where $\Gamma(z)$ is the gamma function.
Using Stirling's formula expressed by 
$\Gamma(z+1) \approx \left(z/e \right)^z
$, \cite{Abramowitz}
we obtain \cite{diMarzio,Shew,ShewYoshikawa} 
\begin{align}
\ln\nu(R)\approx 
\left(\frac{R}{\ell}\right)^{3}
\left[ \left(1-m_p p_i \right) \ln \left(1-m_p p_i \right) - 
\left(1-m_p p_i -\Phi\right) \ln \left(1-m_p p_i -\Phi \right) 
\right] - m_c.
\label{eq:p_5}
\end{align}

In the same way, we rewrite 
\begin{align}
\omega(R)=\prod_{k=0}^{n_i-1} \left(1-k m_p \frac{\ell^3}{R^3}  \right)&=\prod_{k=0}^{n_i-1} 
\frac{m_p \ell^3}{R^3} \left[\frac{R^3}{m_p\ell^3}-k \right] \\
&=\left(\frac{m_p \ell^3}{R^3}\right)^{n_i}
\frac{\Gamma\left[\frac{R^3}{m_p \ell^3}+1\right]}{\Gamma\left[\frac{R^3}{m_p \ell^3}-n_i+1\right]}
. 
\label{eq:p_5_1}
\end{align}
Using Stirling's formula, we obtain 
\begin{align}
\ln \omega(R)\approx -n_i -
\frac{R^3}{m_p \ell^3}
 \left(1-m_p p_i \right) \ln \left(1-m_p p_i \right) .
\label{eq:p_5_2}
\end{align}
From Eq. (\ref{eq:p_5_2}), we find 
\begin{align}
\ln \omega(R)^{m_p} \approx -m_p n_i -
\left(\frac{R}{\ell}\right)^{3}
 \left(1-m_p p_i \right) \ln \left(1-m_p p_i \right) .
\label{eq:p_5_3}
\end{align}
We also applied Stirling's formula to the rest of the factors in Eq. (\ref{eq:p_2}) to obtain 
\begin{align}
\ln \left[\frac{1}{n_i !} \left(\frac{R^3}{\ell^3} \right)^{n_i} \right]
\approx -
\left(\frac{R}{\ell}\right)^{3}
p_i\ln \left( p_i \right) + n_i. 
\label{eq:p_5_4}
\end{align}

The optimal shape of the long polymer  can be obtained by maximizing $\ln W (R)$ with respect to $\alpha$. 
We note that the volume fraction occupied by short polymers can be varied 
even if $n_i$ is not altered when $R$ is changed. 
In this sense, 
$m_p p_i$ depends on the volume fraction of the long polymer, $\Phi$. 
Because $p_i$ depends on $\alpha$ through $\Phi$, it is convenient to define a new variable, 
\begin{align}
r &=\frac{m_p p_i}{\Phi}=\frac{m_p n_i}{m_c} . 
\label{eq:r}
\end{align}
Here, $n_i$ can be regarded as a constant  
for the infinitesimal small variation of $R$. 
In this work, we consider the small variation of $R$ and regard $r$ as being independent of $\alpha$. 

By taking into account Eqs. (\ref{eq:p_5})-(\ref{eq:r}), we obtain 
\begin{multline}
\frac{W (R)}{{R_0}^2} \approx 
\alpha^2 \exp \left[ -\alpha^2 - \frac{m_c^{3/2}}{s} \alpha^3 
\left\{ p_i \ln \left( p_i\right) +
\left(1-\frac{s(1+r)}{\sqrt{m_c}\alpha^3}\right) \ln \left(1-\frac{s(1+r)}{\sqrt{m_c}\alpha^3} 
\right)
\right\}
+\right. \\ \left.
\sqrt{m_c}\,  s\frac{1}{\alpha^3} \chi_{\rm eff} - m_c \left(\chi_{\Phi,s}+r \chi_{p.s} +1+r-\frac{r}{m_p}\right)
\right], 
\label{eq:p_6}
\end{multline}
where the effective $\chi$-parameter is defined by 
\begin{align}
\chi_{\rm eff}&=
\chi_{\Phi,s} -r \Delta \chi +r^2 \chi_{p,s}
\label{eq:chieff}
\end{align}
and $\Delta \chi$ is defined by 
\begin{align}
\Delta \chi= \chi_{\Phi,p}-\chi_{\Phi,s}-\chi_{p,s}. 
\label{eq:Dchi}
\end{align}

The optimal shape of the long polymer  can be obtained by 
%%% revised 
maximizing 
%%% revised 
Eq. (\ref{eq:p_6}) with respect to $\alpha$. 
By calculating $[\alpha^4/2](d \ln W(R))/(d \alpha)=0$ using Eq. (\ref{eq:p_6}), 
we find  
\begin{align}
\alpha^5-\alpha^3 +\frac{3 m_c}{2} \left(1+r-\frac{r}{m_p}\right) \alpha^3+\frac{3}{2} \chi_{\rm eff} \sqrt{m_c} s + 
\frac{3m_c^{3/2}\alpha^6}{2 s} \ln\left[1-\frac{s(1+r)}{\sqrt{m_c}\alpha^3}
\right]=0 . 
\label{eq:p_7}
\end{align}
Equation (\ref{eq:p_7}) was derived previously when $r=0$ using the partition function 
except for a minor difference associated with the definition of $\alpha$. 
\cite{diMarzio,Shew,ShewYoshikawa} 

By further expanding Eq. (\ref{eq:p_7}) in terms of $s/(\sqrt{m_c}\alpha^3)$, 
we obtain 
\begin{align}
\alpha^5-\left(1+\frac{3}{2} \frac{m_c r}{m_p}\right)\alpha^3 -\frac{3s}{4} \sqrt{m_c} \left[(1+r)^2-2\chi_{\rm eff}  \right]-
\frac{s^2}{2\alpha^3}\left(1+r\right)^3=0. 
\label{eq:p_8}
\end{align}

For simplicity, we have expressed the volume of the polymer segment as $\ell^3$. 
In fact, the volume of a polymer segment can be given by $d^2 \ell$
using the diameter of the polymer $d$.
In this case, $s$ is proportional to $d^2 /\ell^2$. 
Below, we regard $s$ as a parameter that characterizes the anisotropy of a polymer segment.

\subsection{Derivation by free energy}
We further show that the partition function given by Eq. (\ref{eq:p_6}) 
 can be derived by combining the mixing entropy
and entropic elastic energy of the long polymer when the mixing entropy is multiplied by the properly defined volume occupied by the long polymer. 

Using the expansion factor $\alpha$, the entropic elastic energy of the long polymer is given by, \cite{Rubinshtein}
\begin{align}
\frac{E_{\rm el}}{k_{\rm B} T}=\alpha^2 -2 \ln \alpha.   
\label{eq:el}
\end{align}

We consider configurations of polymers using a lattice model.  
In the mean field approximation, the mixing entropy at a site inside the long polymer is given by, \cite{Rubinshtein}
\begin{align}
S_{\rm mix} \left(\Phi,p_i\right)= -k_{\rm B} \left[ p_i \ln \left( m_p p_i\right)+ \left(1-\Phi-m_p p_i\right) \ln
\left(1-\Phi - m_p p_i \right) \right]. 
\label{eq:S}
\end{align}
In the absence of the elastic energy given by Eq, (\ref{eq:el}), 
the free energy at a site inside the long polymer can be written as, \cite{Rubinshtein}
\begin{align}
\frac{F_{\rm in}\left(\Phi,p_i\right)}{k_{\rm B} T} &= 
 - \frac{1}{k_{\rm B}} S_{\rm mix} + \chi_{\Phi,p} m_p p_i \Phi +\chi_{\Phi,s} \left(1- m_p p_i -\Phi \right) \Phi +\chi_{p,s} \left(1- m_p p_i -\Phi \right) m_p p_i . 
\label{eq:Fin}
\end{align}

The entropic elastic energy for a long polymer is given in Eq. (\ref{eq:el})  and 
the free energy per site is given by Eq. (\ref{eq:Fin}). 
To add these terms, we should multiply the free energy of a site by a certain volume. 
The real polymer volume can be approximated by $m_c d^2 \ell$. 
We assumed that the effective volume occupied by a long polymer is $4 \pi R^3/(3\ell^3)$. 
By comparing the site free energy given by Eq. (\ref{eq:Fin}) and the probability distribution given by Eq. (\ref{eq:p_6}) 
using the polymer volume given by $4 \pi R^3/(3\ell^3)$, \cite{Vasilevskaya}
the total free energy $F_{\rm in} 4 \pi R^3/(3\ell^3)$ including the elastic energy given by Eq. (\ref{eq:el}) can be expressed as, 
\begin{multline}
\frac{F_{\rm t}\left(\Phi, p_i \right)}{k_{\rm B} T} =\frac{m_c}{\Phi}
\left[p_i \ln \left( m_p p_i\right)+\left(1-\Phi - m_p p_i \right)\ln
\left(1-\Phi - m_p p_i \right) +\chi_{\Phi,p} m_p p_i \Phi + 
\right. \\ \mbox{ } \left. 
\chi_{\Phi,s} \Phi \left(1- m_p p_i -\Phi \right)+\chi_{p,s} \left(1- m_p p_i -\Phi \right)  m_p p_i \right]+ \frac{E_{\rm el}}{k_{\rm B} T}.   
\label{eq:tenergy}
\end{multline}
Equation (\ref{eq:tenergy}) is equal to 
%%% revised 
$-\ln W(R)$ obtained from 
%%% revised 
Eq. (\ref{eq:p_6}) except for the unimportant term $1+r-(r/m_p)$.

In the limit of $p_i=0$, Eq. (\ref{eq:tenergy}) is simplified to the equation obtained before. \cite{Sanchez,Ziv}
The shape equation given by Eq. (\ref{eq:p_7}) can be obtained by minimizing Eq. (\ref{eq:tenergy}) with respect to $\alpha$. 
Previously, the equation representing the osmotic pressure balance was derived from Eq. (\ref{eq:tenergy}) 
and the osmotic pressure balance was imposed 
by regarding the region occupied by the long polymer as being in a phase different from that of the other region. \cite{Vasilevskaya}
We did not impose the osmotic pressure balance at the boundary of the region occupied by the long polymer. 
Instead, we derive the optimal shape by minimization of Eq. (\ref{eq:tenergy}) with respect to the expansion factor. 
In the absence of short polymers, the resultant shape equation simplifies to the conventional form. \cite{Ptitsyn,Post79,Post82}

%%% 
%%%%%%%%%%%%%%%%%%%%%%%%%%%%%%%%%%%%%%%%%%%%%%%%%%%%%%%%%%%%%%%%%%%%%%
\section{The shape equation}
\label{sec:shapeeq}

First, we study the shape equation derived from the partition function analytically.
For convenience, we express Eq. (\ref{eq:p_8}) as  
\begin{align}
f\left(\alpha\right)=\alpha^5 -\left(1+\frac{3}{2} n_i\right) \alpha^3 - \sqrt{m_c}\, c_{\rm t} \tau-
\frac{y}{\alpha^3}=0, 
\label{eq:minimaa}
\end{align}
where parameters are given by, 
\begin{align}
c_{\rm t}&=\frac{3s}{4}, \mbox{ }\tau = (1+r )^2-2 \chi_{\rm eff},
\label{eq:tau}\\
y &= \frac{s^2}{2}
\left(1+r \right)^3,  
\label{eq:y}
\end{align}
and $n_i=m_c r/m_p$ from Eq. (\ref{eq:r}).
When $r=0$, Eq. (\ref{eq:minimaa}) is simplified to the same form of equation as that derived using the virial expansion. 
The term proportional to $\alpha^0$ and $\sqrt{m_c}$ in Eq. (\ref{eq:minimaa}) represents the results of excluded-volume interactions, and 
that proportional to $1/\alpha^3$ represents the results of third-body interactions. \cite{Ptitsyn,Post79,Post82}

Below, we summarize known results for Eq. (\ref{eq:minimaa}) when $n_i=0$ and generalize them for $n_i>0$.
The results can be used when we study the coil--globule transition by solving Eq. (\ref{eq:p_7}) numerically. 

\subsection{Extended state}
Extended states can be characterized by $\alpha >1$. 
If $c_{\rm t}\tau$ is positive, extended states are stabilized according to Eq. (\ref{eq:minimaa}). 
When $m_c \gg 1$, the extended states obey the scaling form 
$\alpha^5 \sim \sqrt{m_c}\, c_{\rm t}\tau$. 
By substituting $s \sim d^2/\ell^2$ and $\alpha \sim R/\left(\sqrt{m_c}\, \ell\right)$
into the scaling form, we obtain the scaling relation 
for $R$ in good solvents, \cite{Rubinshtein,FTanaka}
\begin{align}
R \sim m_c^{3/5} \left(d^2 \ell^3\right)^{1/5} , 
\label{eq:csolvents}
\end{align}
when $\tau > 0$. 
If $\tau=0$, $R \sim m_c^{1/2}$ can be derived, which represents the same scaling form of
the free polymers without taking into account the excluded-volume interactions. 

\subsection{Compact State}
Compact states are characterized by $\alpha \ll 1$. 
When $m_c \gg1$, 
 the compact states obey scaling form, 
$-\sqrt{m_c}\, c_{\rm t}\tau \sim y/\alpha^3$. 
The scaling relation 
for $R$ is obtained as \cite{Rubinshtein,FTanaka}
\begin{align}
R \sim m_c^{1/3} \left(d^2 \ell \right)^{1/3} , 
\label{eq:gsolvents}
\end{align}
when $\tau<0$. 
The scaling of Eq. (\ref{eq:gsolvents}) represents the reasonable assumption  
that the molecular weight of the long polymer is proportional to its volume  given by $R^3$ in the compact state. 

\subsection{Coexistence state}
For the values other than $\alpha \gg 1$ and $\alpha \ll 1$, 
we studied Eq. (\ref{eq:minimaa}) analytically by assuming that $c_{\rm t}\tau$ can take any value and vary independently of $y$. 
The results can be used as a guide for 
numerical solutions  of Eq. (\ref{eq:p_7}), 
where we take into account proper constraints on 
$c_{\rm t} \tau$ and $y$, in particular the fact that they are not independent functions of the concentration of short polymers $p_i$.  

It is convenient to define 
\begin{align}
g(\alpha)=\alpha^5 - \left(1+\frac{3}{2}n_i \right)\alpha^3 -\frac{y}{\alpha^3} . 
\label{eq:galpha}
\end{align} 
Equation (\ref{eq:minimaa}) can be rewritten using $g(\alpha)$ as 
\begin{align}
g(\alpha)=  \sqrt{m_c}\, c_{\rm t} \tau. 
\label{eq:gtau}
\end{align}

If $g(\alpha)$ has a maximum and a minimum in the region of $\alpha>0$, 
the coexistence of coil and globule states is possible depending on the value of $\tau$. 
The above condition can be restated as $h(\alpha)=g'(\alpha)/\alpha^2=0$ has two solutions for $\alpha>0$. 
The explicit condition can be obtained by defining $\alpha^\dagger$, which satisfies
both $h'(\alpha^\dagger)=0$ and $h(\alpha^\dagger)<0$ for $\alpha^\dagger>0$. 
Using $\alpha^\dagger=y^{1/8} 3^{1/4}/5^{1/8}$, \cite{FTanaka}
the condition is given by $0 < y < y^\dagger$, where
\begin{align}
y^\dagger =\frac{3^6}{4^4\times 5^3} \left( 1+\frac{3}{2} n_i\right)^4
= \frac{729}{32000} \left( 1+\frac{3}{2} n_i\right)^4
\approx 0.0228\left( 1+\frac{3}{2} n_i\right)^4.  
\label{eq:yd}
\end{align}
Therefore, the coexistence of coil and globule states may be possible when $0 < y < y^\dagger$ depending on the right-hand side of Eq. (\ref{eq:gtau}).

We should also note that $g(\alpha)<0$ can be transformed into $\alpha^2 -1-(3/2)n_i < y/\alpha^6$. 
If $0<\alpha <1$, {\it i.e.}, $\alpha^2-1<0$, $g(\alpha)<0$ is always satisfied because $y$ is positive according to 
Eq. (\ref{eq:y}). In addition,
$\tau<0$ is required if the solution of $g(\alpha)=\sqrt{m_c} c_{\rm t} \tau$ satisfies $0<\alpha <1$ 
because $g(\alpha)=\sqrt{m_c} c_{\rm t} \tau<0$ and $c_{\rm t}>0$. 
The coexistence of coil and globule states implies that we have two solutions and 
one of them should satisfy $0<\alpha<1$. 
Therefore, coexisting states can be obtained when $\tau<0$. 

%%%%%%%%%%%%%%%%%%%%%%%%%%%%%%%%%%%%%%%%%%%%%%%%%%%%%%%%%%%%%%%%%%%%%%
\section{Shape transition}
\label{sec:Shape transitions} 

We next present numerical results of the shape transitions of the long polymer as a function of $r$.
$r$ is proportional to the number of short polymers 
inside the region occupied by the long polymer $n_i$.

\subsection{Parameter estimation}

%%%%%%%%%%%%%%%%%%%%%%%%%%%%%%%%%%%%%%%
\begin{figure}
\centerline{
\includegraphics[width=0.5\columnwidth]{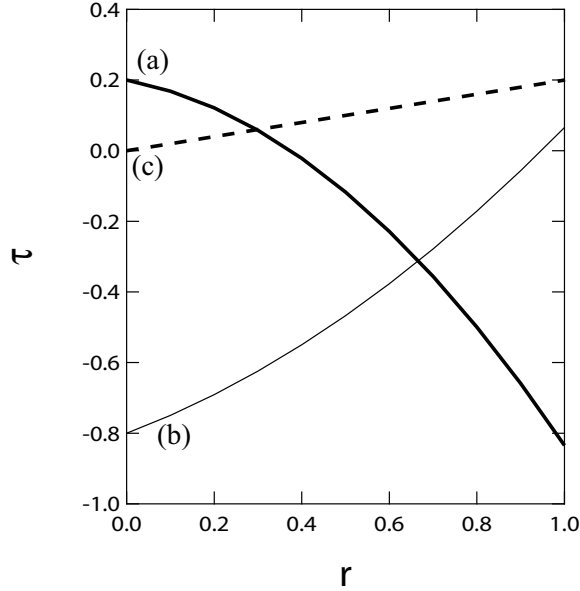}
}
\caption{$\tau$ as a function of $r$. 
$r$ is proportional to the number of short polymers inside the region occupied by the long polymer. 
(a) $\chi_{p,s}=0.9$, $\chi_{\Phi,s}=0.4$, and $\chi_{\Phi,p}=0.18$ (thick solid line). 
(b) $\chi_{p,s}=0.3$, $\chi_{\Phi,s}=0.9$, and $\chi_{\Phi,p}=0.433$ (thin solid line). 
(c) $\chi_{p,s}=0.5$, $\chi_{\Phi,s}=0.5$, and $\chi_{\Phi,p}=0.1$ (dashed line). 
}
\label{fig:tau_CG}
\end{figure}
%%%%%%%%%%%%%%%%%%%%%%%%%%%%%%%%%%%%%%

We first estimate appropriate parameters to discuss the shape transition
by considering $\tau$ as a function of $r$ (Fig. \ref{fig:tau_CG}).
In general, the long polymer is in the elongated state if $\tau>0$ and in the compact state if $\tau<0$,
and the sign of $\tau$ can be changed by changing $r$.
By using Eq. (\ref{eq:chieff}), Eq. (\ref{eq:tau}) can be rewritten as 
\begin{align}
\tau = 1- 2 \chi_{\Phi,s}+2 \left(1+ \Delta \chi \right) r +\left(1-2 \chi_{p,s} \right)r^2.
\label{eq:tau1}
\end{align}
At $r=0$, the elongated coil state can be obtained when $\chi_{\Phi,s}<1/2$,
and the compact globule state can be obtained when $\chi_{\Phi,s}>1/2$.
As for $\chi_{p,s}$, when $\chi_{p,s}>1/2$, $\tau$ becomes negative upon increasing $r$,
as can be understood from Eq. (\ref{eq:tau1}).
In contrast, when $\chi_{p,s}<1/2$, $\tau$ becomes positive upon increasing $r$. 
Based on the above reasoning, if $\chi_{\Phi,s}<1/2$ and $\chi_{p,s}>1/2$ (Fig. \ref{fig:tau_CG} (a)),
the coil-to-globule transition can be obtained,
and if $\chi_{\Phi,s}>1/2$ and $\chi_{p,s}<1/2$ (Fig. \ref{fig:tau_CG} (b)),
the globule-to-coil transition occurs.
Figure \ref{fig:tau_CG} (c) presents the results when $\chi_{\Phi,s}=1/2$ and $\chi_{p,s}=1/2$. 
In this case, $\tau$ slowly increases as $r$ is increased. 
Below, we assume that $m_c=1000$, $m_p=100$, and $s=0.3387$.

\subsection{Results}

\subsubsection{Coil-to-globule transition}

%%%%%%%%%%%%%%%%%%%%%%%%%%%%%%%%%%%%%%%
\begin{figure}
\centerline{
\includegraphics[width=0.5\columnwidth]{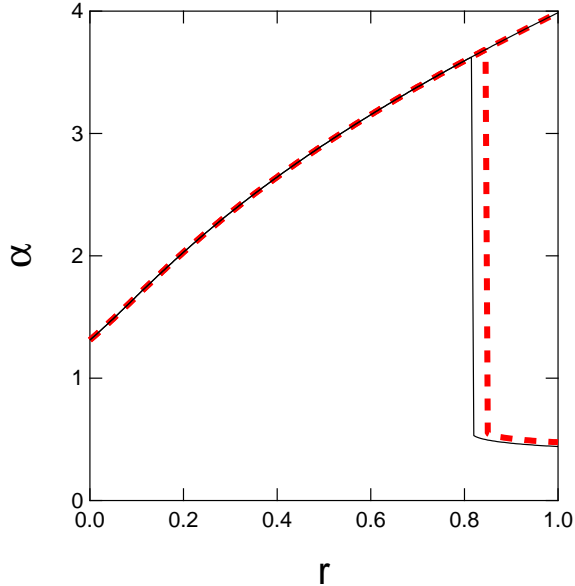}
}
\caption{$\alpha$ as a function of $r$ when $\chi_{p,s}=0.9$, $\chi_{\Phi,s}=0.4$, and $\chi_{\Phi,p}=0.18$. 
Here, $r$ is proportional to the number of short polymers inside the region occupied by the long polymer. 
The red dashed lines were obtained using Eq. (\ref{eq:p_7}) and the solid lines were obtained using 
the approximate expression given by Eq. (\ref{eq:minimaa}) using 
$m_c=10^3$, $m_p=100$, and $s=0.3387$. 
 }
\label{fig:CG_1}
\end{figure}
%%%%%%%%%%%%%%%%%%%%%%%%%%%%%%%%%%%%%%

Figure \ref{fig:CG_1} shows the expansion factor $\alpha$ as a function of $r$
with $\chi_{p,s}>1/2$ and $\chi_{\Phi,s} <1/2$ where the coil-to-globule transition can be expected.
In the absence of short polymers ($r=0$), the long polymer extends in the good solvent characterized
by $\chi_{\Phi,s} <1/2$.
The short polymers tend to aggregate in the poor solvent characterized by $\chi_{p,s}>1/2$.
When there are excluded-volume interactions between the long and short polymers,
the long polymer may shrink if the short polymers tend to aggregate
and the motion of short polymers is hindered by the presence of the long polymer.
Therefore, the globule state can be obtained when $r$ exceeds a certain value.

Figure \ref{fig:CG_1} shows the coil-to-globule transition.
$\alpha$ has two different values that correspond to the coil and globule states
when $r$ exceeds a certain value, and a single value ($\alpha>1$)
when $r$ is less than a certain value.
Such a discontinuous transition suggests that it is a first-order transition
and the compact globule state can be found when $r$ exceeds a certain value.

Incidentally, in Fig. \ref{fig:CG_1}, the dashed line was obtained using the original equation given by
Eq. (\ref{eq:p_7}) and the solid line was obtained using the approximate expression given by Eq. (\ref{eq:minimaa}).
We note that the difference between these lines is small and the qualitative $r$ dependence is well captured
by Eq. (\ref{eq:minimaa}).

\subsubsection{Globule-to-coil transition}

%%%%%%%%%%%%%%%%%%%%%%%%%%%%%%%%%%%%%%%
\begin{figure}
\centerline{
\includegraphics[width=0.5\columnwidth]{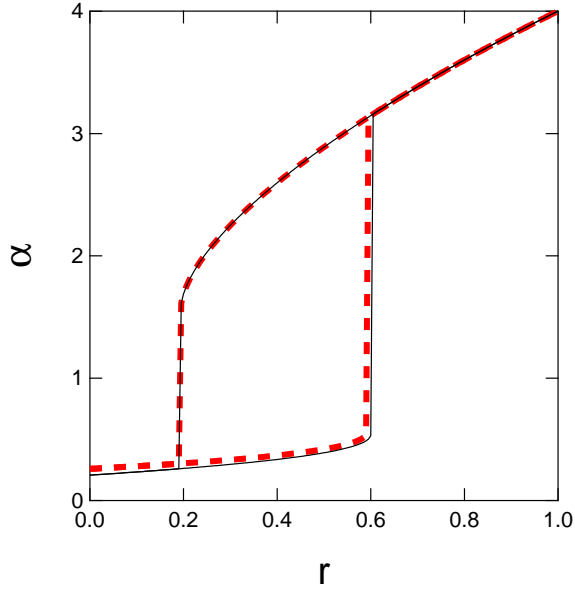}
}
\caption{ 
$\alpha$ as a function of $r$ when $\chi_{p,s}=0.3$, $\chi_{\Phi,s}=0.9$, and $\chi_{\Phi,p}=0.433$. 
$r$ is proportional to the number of short polymers inside the region occupied by the long polymer.
The red dashed lines were obtained using Eq. (\ref{eq:p_7}) and the solid lines were obtained using 
the approximate expression given by Eq. (\ref{eq:minimaa}) using  
$m_c=10^3$, $m_p=100$, and $s=0.3387$. 
}
\label{fig:GC_1}
\end{figure}
%%%%%%%%%%%%%%%%%%%%%%%%%%%%%%%%%%%%%%

Figure \ref{fig:GC_1} shows the results for $\chi_{p,s}<1/2$ and $\chi_{\Phi,s} >1/2$ where
the globule-to-coil transition can be expected.
In the absence of short polymers ($r=0$),
the long polymer is in a compact state in a poor solvent characterized by $\chi_{\Phi,s}>1/2$. 
In contrast, the short polymers are in good solvents.
If a sufficient number of short polymers tend to be dispersed and involved in excluded-volume interactions with the long polymer, 
the extended coil state is expected. 
Figure \ref{fig:GC_1} clearly shows that the long polymer changes from the compact globule state to
the extended coil one as $r$ increases.
The transition is discontinuous.  
It should be noted that the extended coil state can be expected if a sufficient number of short polymers 
are involved in excluded-volume interactions with the long polymer regardless of the values of $\chi_{p,s}$ and $\chi_{\Phi,p}$. 
Indeed, we obtain a similar figure even when the short polymers can be regarded as a poor solvent
for the long polymer  characterized by $\chi_{\Phi,p} =0.6>1/2$ if the other values in Fig. \ref{fig:GC_1} are fixed. 
The discontinuous transition to the extended coil state is also obtained
when $\chi_{p,s}$ is changed to $\chi_{p,s}=0.6$ and the other values in Fig. \ref{fig:GC_1} are fixed.
In the latter two cases, 
the transition to the extended state is mainly caused by the excluded-volume interactions between the long and short polymers.

\subsubsection{Continuous coil expansion}

%%%%%%%%%%%%%%%%%%%%%%%%%%%%%%%%%%%%%%%
\begin{figure}
\centerline{
\includegraphics[width=0.5\columnwidth]{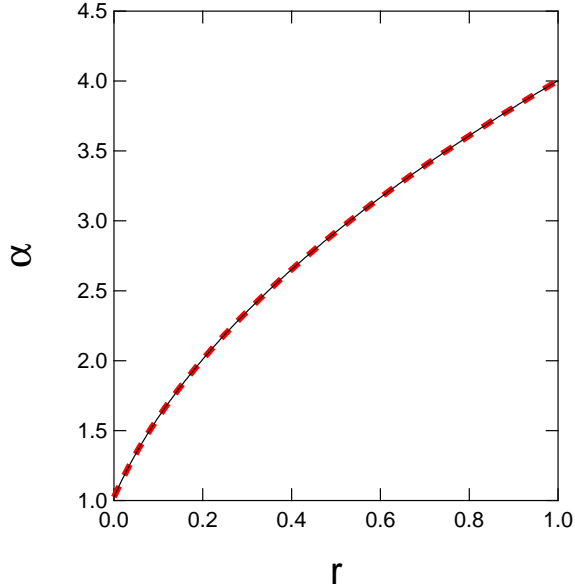}
}
\caption{$\alpha$ as a function of $r$ when $\chi_{p,s}=0.5$, $\chi_{\Phi,s}=0.5$, and $\chi_{\Phi,p}=0.1$. 
$r$ is proportional to the number of short polymers inside the region occupied by the long polymer. 
The red dashed line was obtained using Eq. (\ref{eq:p_7}) and the solid line was obtained using 
the approximate expression given by Eq. (\ref{eq:minimaa}) using 
$m_c=10^3$, $m_p=100$, and $s=0.3387$. 
}
\label{fig:CG_2nd}
\end{figure}
%%%%%%%%%%%%%%%%%%%%%%%%%%%%%%%%%%%%%%

Figure \ref{fig:CG_2nd} shows the results for $\chi_{p,s}=0.5$, $\chi_{\Phi,s} =0.5$ and $\chi_{\Phi,p}=0.1$. 
The extended coil state expands gradually and the discontinuous transition is not obtained.

\subsection{Discussion}
In experiments, coil-to-globule transitions of DNA were observed when the solvent acted as a good solvent for both long DNA and 
short polyethylene glycol (PEG). \cite{Maeda,Vasilevskaya}
 Although a coil-to-globule transition can be obtained according to our theoretical approach 
 when the solvent acts as a good solvent for the long DNA, 
 the condition for short polymers is different. 
By including the translational entropy of small ions, 
experimental results have been reproduced using mean field theory. \cite{Vasilevskaya} 
Although such an ion effect is ignored in our theory, we systematically varied Flory's $\chi$-parameters and 
obtained the general conditions for transitions between coil and globule states for neutral polymers. 
In some experiments, a coil--globule--coil transition was observed. \cite{Vasilevskaya}
Further study is needed to understand theoretically the globule-to-coil reentrance transition.

%%%%%%%%%%%%%%%%%%%%%%%%%%%%%%%%%%%%Comment on the results
 %%%%%%%%%%%%%%%%%%%%%%%%%%%%%%%%%%%%%%%%%%%%%%%%%%%%%%%%%%%%%%%%%%%%%%
\section{Equilibrium between outside and inside the area occupied by a long polymer}
\label{sec:equilibrium shape} 

Here we discuss a way to determine the number of short polymers $n_i$ inside the region
occupied by the long polymer.
We consider the situation where short polymers can enter the region occupied by the long polymer
and the concentration of short polymers inside the region occupied by the long polymer is different
from that in the region outside the long polymer.
The concentrations of short polymers outside and inside the region occupied by the long polymer
can be expressed as $p/\ell^3$ and $p_i/\ell^3$, respectively.
The relationship between the concentrations of short polymers in these regions can be determined by imposing the equality of
chemical potentials.
The free energy at a site outside the region occupied by the long polymer  
can be given by, \cite{Rubinshtein}
\begin{multline}
\frac{F_{\rm out}}{k_{\rm B} T} = 
 p \ln \left( m_p p\right)+ \left(1-m_p p-m_c c \right) \ln
\left(1 - m_p p -m_c c \right)+ c \ln \left( m_c c\right) + 
\\
\mbox{ }\chi_{p,s} m_p p \left(1 - m_p p -m_c c \right) + \chi_{\Phi,s} m_c c \left(1 - m_p p -m_c c \right)+
\chi_{\Phi,p} m_c c\, m_p p 
\label{eq:Fout}
\end{multline}
in the mean field approximation, where $c$ denotes the concentration of the long  polymers
outside the region occupied by a single long polymer.
When short polymers can move across the boundary of the region occupied by a long polymer, 
 the chemical potential outside the region containing the long polymer, $\mu=\partial F_{\rm out}/\partial p$, should be equal to that inside the area with the long polymer,  $\partial F_{\rm in}/\partial p_i$, which was determined using Eqs. (\ref{eq:S})-(\ref{eq:Fin}).
The result can be written as, 
 \begin{align}
 \ln \left(p/p_i \right) -m_p \ln \left[ \frac{1-m_c c - m_p p }{1-\Phi - m_p p_i } \right] = \Delta \chi 
 m_p\left( \Phi-m_c c \right)+ 2 \chi_{p,s} m_p^2 \left(p-p_i \right), 
 \label{eq: chemeq}
 \end{align}
 where $\Delta \chi$ is defined in Eq. (\ref{eq:Dchi}). 
 Equation (\ref{eq: chemeq}) can be expressed as 
 \begin{multline}
 \frac{m_p p}{
 \left( 1 - m_c c - m_p p \right)^{m_p}}
 \exp \left[ 
 \Delta \chi m_p m_c c - 2 \chi_{p,s} m_p^2 p \right] = \\
\mbox{ }  \frac{m_p p_i}{
 \left( 1 - \Phi - m_p p_i \right)^{m_p}}
 \exp \left[ 
 \Delta \chi m_p \Phi - 2 \chi_{p,s} m_p^2 p_i \right] . 
 \label{eq:chemeqPhi}
 \end{multline}
$c$ and $p$ for the regions outside 
and $\Phi$ and $p_i$ for those 
inside the area occupied by a specific long polymer are related by 
Eq. (\ref{eq:chemeqPhi}).  

It is more convenient to rewrite Eq. (\ref{eq:chemeqPhi}) in terms of $r$ and $\alpha$, 
where $p_i$ is replaced by $r$ using Eq. (\ref{eq:r}),  
$m_p p_i = r \Phi$. 
Equation (\ref{eq:chemeqPhi}) can therefore be expressed as 
\begin{align} 
\frac{r s}{\sqrt{m_c}\alpha^3}\frac{\exp \left[ -\Delta \chi m_p  m_c c - \left( 2\chi_{p,s} r- \Delta \chi \right)m_p \frac{s}{\sqrt{m_c}\alpha^3}
\right]}{\left(1-(1+r)s/\left(\alpha^3 \sqrt{m_c} \right) \right)^{m_p}} =\frac{m_p p\exp\left(-2 \chi_{p,s} m_p^2 p \right)}{\left(1- m_p p -m_c c \right)^{m_p}} 
. 
\label{eq: chemeq2_1}
\end{align}
The optimal expansion factor, $\alpha$, can be determined from Eq. (\ref{eq:p_7}) and Eq. (\ref{eq: chemeq2_1}). 

\begin{figure}
\centerline{
\includegraphics[width=0.5\columnwidth]{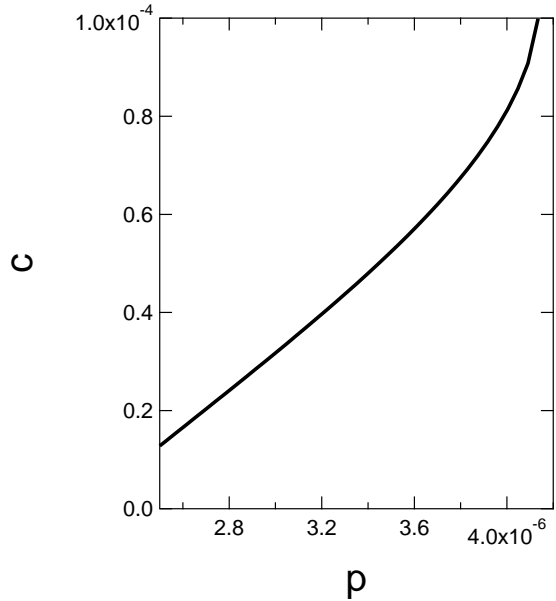}
}
\caption{$c$ as a function of $p$ obtained using Eq. (\ref{eq: chemeq2_1}) when $r=0.5$ and $\alpha=2.90896$, with
$\chi_{p,s}=0.9$, $\chi_{\Phi,s}=0.4$, $\chi_{\Phi,p}=0.18$,  
$m_c=10^3$, $m_p=100$, and $s=0.3387$. 
}
\label{fig:GC_p_s}
\end{figure}

In Fig. \ref{fig:GC_p_s}, 
we show $c$ as a function of $p$ for the parameters used to draw Fig. \ref{fig:CG_1}. 
The results indicate that the same values of $\alpha$ and $r$ can be obtained if $c$ and $p$ satisfy the relationship represented by the line in  Fig. \ref{fig:GC_p_s}. 
%%%%%%%%%%%%%%%%%%%%%%%%%%%%%%%%%%%%%%%
%%%%%%%%%%%%%%%%%%%%%%%%%%%%%%%%%%%%%%

%%%%%%%%%%%%%%%%%%%%%%%%%%%%%%%%%%%%%%%%%%%%%%%%%%%%%%%%%%%%%%%%%%%%%%
\section{Conclusion}
\label{sec:Discussion}
%%%%%%%%%%%%%%%%%%%%%%%%%%%%%%%%%%%%
We theoretically investigated the excluded-volume and interaction effects of
added polymer solvents on the transition between the coil and globule states of a long polymer. 
The excluded-volume effect was taken into account by generalizing the partition function
of the long polymer to include the excluded-volume effect of short polymers.
The interactions between the segments of short polymers, those of long polymer and solvents
were described using Flory's $\chi$-parameters as in other theories. 
Following this approach, we obtained the optimal expansion factors denoted by $\alpha$ from the maxima of the partition function.
To be more precise, the partition function was maximized by assuming that the number of short polymers
$n_i$ inside the region occupied by the long polymer was conserved
against the infinitesimal change in the gyration radius $R$ of the long polymer.
The shape equation was given in terms of the effective Flory's $\chi$-parameter denoted by $\chi_{\rm eff}$.  
$\chi_{\rm eff}$ was expressed by $n_i$ as well as 
Flory's intrinsic $\chi$-parameters.

The shape equation was simplified using asymptotic expansion
in terms of $\alpha$.
In the absence of short polymers, the simplified equation reduces to the classical expression derived by the virial expansion.
%The shape transition of the long polymer has been studied using the shape equation 
%by taking into account the excluded-volume effect by short polymers. 
The simplified shape equation captured the main features of the nonlinear shape equation and 
thus is very useful to analytically investigate the shape transition of polymers.
Previously, the shape transition had been extensively studied using the simplified equation obtained
by the virial expansion in the absence of short polymers. \cite{FTanaka}
Guided by this simplified equation, we systematically studied the shape transition of a long polymer induced by the presence of short polymers.

Using the simplified shape equation and the effective $\chi$-parameter $\chi_{\rm eff}$, 
we showed that coil-to-globule transitions can be induced by increasing $n_i$ 
when the solvent acts as a good solvent for the long polymer and a poor solvent for
the short polymers.  
The long polymer tends to extend in a good solvent and short polymers tend to aggregate in a poor solvent. 
Under the competitive effect induced by such a solvent, 
the long polymer shrinks if the long polymer segments are involved in excluded-volume interactions
with a sufficient number of short polymers.
These transitions were confirmed by numerical calculations using the complete shape equation.
In contrast, 
globule-to-coil transitions can be induced by increasing $n_i$ when the solvent acts as a poor solvent for the long polymer. 
This is because long polymer shrinks in the poor solvent, but expands through the excluded-volume interaction between the long polymer segments and short polymers.
The globule-to-coil transition occurs at a lower concentration of short polymers when 
the solvent acts as a good solvent for the short polymers compared to that in the case where 
the solvent acts as a poor solvent for the short polymers. 
The globule-to-coil transition of a long polymer results from 
excluded-volume interaction with a sufficient number of short polymers 
and the threshold value for the number of short polymers can be low when 
the short polymers tend to be well dispersed in the good solvent. 

The excluded-volume effects can be also taken into account by considering the site-mixing entropy of the lattice model. 
However, there was some ambiguity in defining the effective volume of a long polymer to combine 
the site-mixing entropy and entropic elastic energy of a long polymer. 
We show that the minimization of the total free energy leads to the equation for the optimal expansion factor 
derived using the partition function 
if the mean volume is defined by 
$4 \pi R^3/(3 \ell^3)$, 
where $\ell$ is the Kuhn length and $R$ denotes 
the radius of gyration of the long polymer. 

%%% revised 
In a related topic, 
effects of macromolecular crowding on protein folding has been studied both experimentally and
theoretically \cite{HXZhou08,HXZhou04,Zhou,Ellis,Berg}. 
%A major interaction in crowded environments is cause by the excluded volume effect.
Protein folding stability was studied 
using  a reduction factor for chain conformations by crowding objects. \cite{Zhou,HXZhou08,HXZhou04}
The reduction factor for folding states was obtained from a scaled-particle theory.\cite{Bagchibook,Lebowitz,Stillinger}
The reduction factor for unfolding states was expressed using 
the unoccupied volume to place the protein 
and the fraction    
of the reduced protein conformations due to prohibiting overlapping with crowding objects. 
By calculating the folding free energy using a reduction factor,  
%and comparing the difference between the folding free energy of folded and unfolded states,  
the folding stability was discussed. 
Although the method of calculation is different, 
entropy reduction by excluded volume effect was taken into account 
both in the present theory and in the theory of protein folding stability.  
In this sense, this study gives another insight into the effect of macromolecular crowding 
using a single polymer chain. 
%shows another theoretical approach to treating polymer folding in the macromolecular crowding
%environment using the partition function.

In this paper, we have implicitly assumed that the long polymer is flexible. 
%%% revised 
Flexibility of conjugated polymers can be increased by formation of defects. 
Although conjugated polymers are stiff, 
various compact shapes of conjugated polymers were simulated by introducing defects. \cite{Hu}
%%% revised
It has been pointed out that the folding characteristics of a long semiflexible chain 
differ from those of a short stiff chain. \cite{Sakaue}
Recently, it was demonstrated that a polymer of intermediate length formed various types of collapsed states,
while the spherical collapsed state was obtained for long polymers. \cite{Bagchi}
Spherical symmetric globules can be also classified into several states. \cite{Nogchi96,Nogchi,Montesi,Schnurr}
It was also shown that 
combined effect of rigidity and 
hydrophobicity is correlated with the 
folding kinetics and structures.\cite{Srinivas,Srinivas03}
To identify the shape of a long polymer, its stiffness, 
chemical interactions and finite length should be explicitly considered. 
Although such complexity was ignored in this paper,
the competitive and cooperative effects of excluded volume and the interaction
between monomers and solvents were taken into account in this work using the partition function
in the mean field approximation.
By maximizing the partition function, we showed that various kinds of transition between
the coil and globule states are possible by adding two different solvents,
even if one of the solvents is short polymers.
%%% revised 
In particular, 
we qualitatively show that the shape transition can be induced by the excluded volume interactions 
if solvent acts as good (poor) solvent for 
a long polymer and poor (good) solvent for short macromolecules. 
The mechanism will apply for real polymers having stiffness and other chemical interactions. 
%The quantitative verification of the mechanism can be performed by simulations and experiments.

%%%%%%%%%%%%%%%%%%%%%%%%%%%%%%%%%%%
\acknowledgments
KO was financially supported by a Kenkyu-Josei Grant from Senshu University.

%KS was supported by a Grant-in-Aid for Scientific Research
%(grant No.\ 24540439) from MEXT  of Japan.

%%%%%%%%%%%%%%%%%%%%%%%%%%%%%%%%%%%
%%%%%%%%%%%%%%%%%%%%%%%%%%%%%%%%%%%
\appendix 
%%%%%%%%%%%%%%%%%%%%%%%%%%%%%%%%%%%
\section{Derivation of Eq. (\ref{eq:p_2})}
\label{sec:AppA}
%%% revised
If the excluded volume interactions are ignored, 
the total number of configurations of a long chain 
is proportional to $R^2 \exp \left( -  \frac{3R^2}{2 m_c\, \ell^2} \right)$.
In the presence of excluded volume interactions, the number of configurations is influenced by 
the presence of short polymers. 

We first consider the possible ways to place short polymers under excluded volume effect. 
%%% revised
When the first segment of the $k+1$th short polymer is placed inside the region occupied by the long polymer, 
the number of possible ways to place the first segment is proportional to the free volume 
$(R^3/\ell^3)-k m_p$, where $k m_p$ denotes the total segments previously placed in the region. 
The second segment of the $k+1$th short polymer can be placed to the nearest void if the space is allowed. 
When $k$ short polymers have been previously placed randomly inside the area occupied by the long polymer, 
the probability of finding a sufficient size of nearest void is given by $1-k m_p (\ell^3/R^3)$ in the mean field approximation. 
The probability of adding $m_p-1$ segments of the $k+1$th short polymer 
after placing the first segment can be given by $[1-k m_p (\ell^3/R^3)]^{m_p-1}$.
In this way, the number of distinct configurations to place 
all $n_i$ short polymers inside the region occupied by a long polymer is proportional to 
the factor 
\begin{align}
\frac{1}{n_i! }
\left[ \prod_{k=0}^{n_i-1} \left(\frac{R^3}{\ell^3}\right) \left(1-k m_p \frac{\ell^3}{R^3}  \right)^{m_p} \right],
\label{eq:app_1}
\end{align}
where $ 1/n_i!$ is multiplied to account for 
$n_i!$ ways of ordering indistinguishable short polymers. 

%%% revised
Now, we consider the excluded volume effect of the long polymer segments under the presence of short polymers. 
For each configuration whose incidence is proportional to $R^2 \exp \left( -  \frac{3R^2}{2 m_c\, \ell^2} \right)$, 
it requires that all $m_c$ successive segments of the long polymer can be placed in a void. 
The probability of appearance of such a void can be estimated as follows. 
When short polymers and 
$j$ segments of the long polymer are already placed, 
%%% revised
the next segment of the long polymer can be placed in the nearest void and the probability of finding the void is given by the volume fraction expressed by 
$1- m_p p_i - j (\ell^3/R^3)$ in the mean field approximation. 
The probability of placing $m_c$ segments of the long polymer 
%%% revised
in a void 
%%% revised
can be given by 
$\prod_{j=0}^{m_c-1}\left[1- m_p p_i - j  (\ell^3/R^3) \right]$.
By taking into account all the above factors, 
we obtain 
\begin{align}
\frac{1}{n_i! }
\left[ \prod_{k=0}^{n_i-1} \left(\frac{R^3}{\ell^3}\right) \left(1-k m_p \frac{\ell^3}{R^3}  \right)^{m_p} \right]
\left[ \prod_{j=0}^{m_c-1} \left( 1- m_p p_i - j  \frac{\ell^3}{R^3} 
\right)
\right] .
\end{align}
By also taking into account the interaction energies and expansion potential of the long polymer, 
we obtain  Eq. (\ref{eq:p_2}).

%%%%%%%%%%%%%%%%%%%%%%%%%%%%%%%%%%%

\end{document}